\documentclass[12pt]{article}
\usepackage{epsfig}
\usepackage{persdefs}
\newlength{\dinwidth}
\newlength{\dinmargin}
\newlength{\figwidth}
\newcommand\gH[1]{\gammh{H}{#1}}
\newcommand\gK[1]{\gammh{K}{#1}}
\newcommand\gstarH[1]{\gstar{H}{#1}}
\newcommand\gstarK[1]{\gstar{K}{#1}}

\newcommand\intph{\int\frac{dp'}{(2\pi)^d}}
\newcommand\intpc{\int_{\Bc}\frac{dp}{(2\pi)^d}}
\newcommand\intpsl{\int_{\Bc}\frac{dp}{N_f\,(2\pi)^d}}
\newcommand\suml{\sum_{l\in\Z^d}}

\newcommand\pl{p \!+\! 2\pi l}

\newcommand\Hl{\Hh(l)}

\newcommand\epsKtext{\prod_{\nu\in K}             (-1)^{l_\nu}}
\newcommand\psih{ \hat{\psi}}
\newcommand\psibh{\hat{\psib}}

\newcommand\varphib{\bar{\varphi}}

\newcommand\phixH{  \varphi (x,H) }

\renewcommand{\Z}{Z \!\!\! Z}
\renewcommand{\R}{{\kern+.25em\sf{R}\kern-.78em\sf{I} \kern+.78em\kern-.25em}}
\makeatletter
\@addtoreset{equation}{section}
\makeatother

\begin{document}

\vspace*{1cm}
\begin{center}
{\Large Truncated Perfect Actions \\
\ \\
for Staggered Fermions}

\vspace*{1cm}

W. Bietenholz$^{\rm ~a}$ and H. Dilger$^{\rm ~b}$
\vspace*{7mm}

$^{\rm a}$ HLRZ c/o Forschungszentrum J\"{u}lich \\
D-52425 J\"{u}lich, Germany \\

\vspace*{7mm}

$^{\rm b}$ Institute for Theoretical Physics I \\
WWU  M\"{u}nster \\
Wilhelm-Klemm Str. 9 \\
D-48149 M\"{u}nster, Germany

\vspace*{1cm}

Preprint HLRZ 1998-10

\end{center}

\vspace*{1cm}

We discuss the behavior of free perfect staggered fermions
and truncated versions thereof. The study includes
flavor non-degenerate masses. We suggest a new blocking scheme,
which provides excellent locality of the perfect lattice action.
A truncation procedure adequate for the structure of staggered
fermions is applied. We consider spectral and thermodynamic
properties and compare truncated perfect actions, Symanzik
improved and standard staggered fermions in two and four
dimensions.

\newpage

%%%%%%%%%%%%%%%%%%%%%%%%%%%%%%%%%%%%%%%%%%%%%%%%%%%%%%%%%%%%%%%%%%%%%%%%%%%
\section{Introduction}

The inclusion of dynamical fermions in simulations of
gauge theories, in particular QCD, is a major issue in lattice
field theory. Such simulations have
generally been performed either using Wilson fermions \cite{Wilfer}
or staggered fermions \cite{KS}. The latter formulation
is especially useful in the chiral limit, because the remnant chiral
symmetry $U(1) \otimes U(1)$ protects the zero fermion mass
from renormalization. As a related virtue, its artifacts due
to the lattice spacing $a$ are only of $O(a^{2})$, whereas
they are of $O(a)$ for Wilson fermions interacting by gauge fields.

It is now widely accepted that
the above lattice actions should be improved, so that
the lattice spacing artifacts are suppressed and coarser
lattices can be used, particularly in QCD simulations \cite{LAT97}.
There are essentially two improvement strategies in the literature.
In Symanzik's program \cite{Sym} the action is improved
order by order in $a$. For QCD with Wilson fermions
this has been realized on-shell to the first order on the
classical level \cite{SW} and recently also on the
quantum level \cite{Alpha}. Less work has been devoted
to the improvement of staggered fermions, perhaps because
the artifacts in the standard formulation are already smaller.
However, S. Naik has applied Symanzik's program on-shell, where he
improved the free staggered fermion by adding more couplings
along the axes \cite{Naik}, and
the Bielefeld group did the same by adding diagonal couplings
\cite{Bielef}. 
Furthermore, the MILC collaboration achieved a 
reduced pion mass by treating the gauge variable as a 
``fat link'' \cite{MILC}.
Finally, some work on improved operators has been done in
this framework \cite{oper}.

The other promising improvement scheme is non-perturbative
in $a$ and uses renormalization group concepts. 
It has been known for a long time that there are perfect
lattice actions in the parameter space, i.e.\ actions without
any lattice artifacts \cite{WiKo}. Recently it has been
suggested to approximate them for asymptotically free theories
as ``classically perfect actions'' \cite{HN}, which works very 
well in a number of two dimensional models \cite{HN,GN,Toy}.

For free or perturbatively interacting fields, perfect actions
can be constructed analytically in momentum space. For Wilson
type fermions this has been carried out to the first order in the
gauge coupling in the Schwinger model \cite{Schwing} and in QCD 
\cite{QuaGlu}. A technique called ``blocking from 
the continuum'' was extremely useful for this purpose. One expresses
all quantities in lattice units after the blocking, and sends the 
blocking factor to infinity. Hence the blocking process starts from
a continuum theory, and it does not need to be iterated in order
to identify a perfect action.

For staggered fermions a block variable renormalization group
transformation (RGT), which does not
mix the pseudo flavors and which does therefore preserve the
important symmetries, has been suggested in Ref. \cite{KMS}.
It requires an odd blocking factor $n$. Iterating the $n=3$
block variable RGT, a fixed point action, i.e.\ a perfect action
at infinite correlation length, has been constructed 
\cite{GN,Dallas}. Also for staggered fermions blocking from the
continuum is applicable \cite{MaiMack}. 
This has been carried out for a general (flavor non-degenerate)
mass term,
revealing the intimate relation to the Dirac-K\"{a}hler fermion 
formulation in the continuum \cite{Dilg},
and also including a suitable treatment of the gauge field
\cite{BBCW}.
Using the generalization to a flavor non-degenerate mass, the spectral 
doublers inherent to the staggered fermion formulation might be treated
as physical flavors in a QCD simulation.

For any blocking scheme, the perfect action displays the continuum
scaling behavior by definition.
In practice, however, the approximations in the determination of
an applicable quasi perfect action violate this property to some
extent. With this respect, the choice of the RGT is very important.

In the present paper, we first review the procedure of blocking
staggered fermions from the continuum, adding
new aspects. We then discuss the optimization of locality,
in the sense of an extremely fast exponential decay of the
couplings in coordinate space. This property
is crucial for practical applications, because we have to 
truncate the couplings to a small number, which is manageable in
simulations. 
%Of course the truncation violates the perfectness,
%but for excellent locality this violation
For excellent locality, the violation of perfectness due to truncation
is not too harmful. Moreover, the virtues of staggered fermions
come into play: the truncation induces errors of $O(a^{2})$
at most, and the remnant axial symmetry is preserved in the
truncated perfect action, hence it still excludes
additive mass renormalization. On the other hand,
a strong mass renormalization is a severe problem
for truncated perfect fermions of the Wilson type \cite{LAT96,TdG}.

We suggest
a new blocking scheme, which leads to a higher degree of locality
than the usual block average method. Still the couplings decay
exponentially in coordinate space. We then truncate them 
to a short range by
means of mixed periodic and antiperiodic boundary conditions,
which are particularly suitable for staggered fermions.
We finally compare the spectra and the scaling of thermodynamic
quantities for several truncated perfect, Symanzik improved
and standard staggered fermions. The discussion of the
spectrum includes also the case of non-degenerate masses of
the pseudo flavors. It turns out that the new blocking scheme,
which we call ``partial decimation'', does consistently yield
excellent results. Moreover, for that scheme,
optimal locality is reached by the simple $\delta$ function RGT
in the massless case, in contrast to
the block average scheme.

Simulation results for the Schwinger model based on this action
will be presented in a subsequent paper \cite{prep}.

%%%%%%%%%%%%%%%%%%%%%%%%%%%%%%%%%%%%%%%%%%%%%%%%%%%%%%%%%%%%%%%%%%%%%%%%%%%
\section{The Staggered Blocking Scheme} \label{Scheme}

The construction of perfect actions for free staggered fermions by blocking
from the continuum has been described in Refs. \cite{Dilg,BBCW}. 
Let us briefly review this procedure. 
The staggered blockspin fermions are defined in two steps. First we transform
the $N_f=2^{d/2}$ flavors of continuum Dirac spinors 
$\psi_a^b(x)$ ($a$: spinor
index, $b$: flavor index) into the Dirac-K\"ahler (DK) representation by
$\phixH$.%
\footnote{For the relation of the DK formulation of continuum 
fermions \cite{EK62} with staggered lattice fermions we 
refer to Ref. \cite{BJ}. A relation to the
present block spin transformation is discussed in Ref. \cite{Dilg}.}
These functions are considered as component functions of inhomogeneous
differential forms
\eqa \label{components}
    \Phi \ = \ \sum_H \phixH \, dx^H \ , \ \ 
    dx^H \ = \ dx^{\mu_1} \wedge \dots \wedge dx^{\mu_h} \ ,
\eqb
where $H=\{\mu_1,\dots,\mu_h\},\ \mu_1 < \dots < \mu_h $ is a multi-index.
Transformation and inverse transformation read
\Eqa
        \phixH & = & \frac{1}{\sqrt{N_f}} \sum_{ab} \gstarH{ab} \, 
                       \psi_a^b(x) \ , \ \ \ 
      \gamma^H \ = \ \gamma^{\mu_1} \gamma^{\mu_2} \dots \gamma^{\mu_h} \ ,
                                                          \label{psitophi}\\
   \psi_a^b(x) & = & \frac{1}{\sqrt{N_f}} \sum_H \gH{ab} \, \phixH \ .
                                                            \label{phitopsi}  
\Eqb

Second, we introduce a coarse lattice of unit spacing
$\Gammab=\{\yb \,|\, \yb_\mu=\nb_\mu\}$, which is a sublattice of
$\Gamma =\{ y  \,|\,  y_\mu=n_\mu/2\}$, with $\nb_\mu,n_\mu\in\Z$. The fine
lattice points $y$ are uniquely decomposed as ($\muh$ is the unit vector in 
$\mu$--direction)
\eqa
    y \ = \ \yb \, + \, e_H/2 \ , \ \ e_H \ = \ \sum_{\mu\in H} \muh \ .
\eqb 
Thus the multi-index $H(y)$ defines the position of a fine lattice 
point $y$ with respect to the coarse lattice $\Gammab$.
Now the blockspin variables $\phi(y)$ can be defined as averages of the
component functions $\varphi(x,H(y))$, 
with a normalized weight $\Pi(x\!-\!y)$,
$\int\!dx\,\Pi(x\!-\!y)=1$, which is assumed to be even and
peaked around $x=y$,
\eqa \label{blockspin}
    \phi(y) \ = \ \frac{1}{\sqrt{N_f}} \sum_{ab}   
                  \gstar{H(y)}{ab} \, \int dx \ \Pi(x-y) \ \psi_a^b(x) \ .
\eqb
This scheme has been proposed first in Ref. \cite{MaiMack}. 
Its peculiarity is that
the staggered block centers depend on the multi-index $H$ of the 
Dirac-K\"ahler component functions $\phixH$. 
Block average (BA) means in this case average over the overlapping lattice
hypercubes $[y]=\{x | -1/2 \leq (x_\mu-y_\mu ) \leq 1/2 \}$. 
This scheme is given by
$\Pi = \Pi_{BA}$, $\Pi_{BA}(x) = 1$ for $x\in[y]$, $\Pi_{BA}(x) = 0$
otherwise.

For the following calculation we diagonalize the lattice action using the
staggered symmetries. Here it is important that fine lattice shifts are no
symmetry transformations. However, combination with site-dependent sign 
factors gives rise to the non-commuting flavor symmetry transformations
\cite{symmetry}.
Therefore we replace ordinary Fourier transformation by harmonic analysis
with respect to flavor transformations and coarse lattice translations. 
We thus obtain a modified momentum representation which intertwines Fourier
transformation and the transition back from DK fermions to the Dirac basis,
\begin{eqnarray}
   \phi_a^b(p) & = & \sum_y \, e^{ipy} \, \gammh{H(y)}{ab} \, \phi(y)
 \ , \nonumber \\
\bar \phi_a^b(p) & = & \sum_y \, e^{ipy} \, \gammh{H(y)}{ab} \, 
\bar \phi(y) \ , \qquad  p \in \Bc \ = \ ]-\pi ,\pi ]^{d} \ .
\label{ytopab}
\end{eqnarray}
Inserting \eqref{blockspin} we find
\eqa
    \phi_a^b(p) \ = \ \frac{1}{\sqrt{N_f}} \intph \sum_{a'b'} \ \Pi(p') \ 
          \psi_{a'}^{b'}(p') \ \sum_y \ e^{i(p-p')y} \ 
          \gammh{H(y)}{ab} \gstar{H(y)}{a'b'} \ ,     \label{direct}
\eqb
where $\psi_a^b(p), \Pi(p)$ denote the Fourier transform of $\psi_a^b(x),
\Pi(x)$. The last sum can be re-written as
\Eqa
    \sum_{\yb\in\Gammab}\,e^{iq\yb}\ \sum_K e^{iqe_K/2} \ \gK{ab}\gstarK{a'b'} 
    & = & (2\pi)^d \suml
\delta(q-2\pi l)\ \sum_K \prod_{\mu\in K} (-1)^{l_\mu}
          \ \gK{ab} \gstarK{a'b'} \nonumber\\
    & = & N_f \, (2\pi)^d \suml \delta(q-2\pi l) \ \gammh{\Hl}{aa'} 
                                   \gdagg{\Hl}{b'b} \ . 
\Eqb
We have used the orthogonality of the $\gamma$--matrix elements
\eqa \label{sumH}
    \sum_H \gH{ab} \, \gstarH{a'b'} \ = \ N_f \ \delta_{aa'} \,\delta_{bb'}\ ,
\eqb
and $\Hl$ is defined by $H(l) = \{ \mu \,|\,  l_\mu \,\mbox{is odd} \}$,
$\Hh=H$ for $h$ even, $\Hh=\{\mu\,|\,\mu\not\in H\}$ for $h$ odd.
Finally \eqref{direct} becomes (summation over double spin and flavor indices
is understood)
\eqa \label{direct-result}
  \phi_a^b(p) = \sqrt{N_f} \ \suml \ \Pi(\pl) \ \psih_{a'}^{b'}(\pl)\ ,\ 
  \psih_a^b(\pl) = \gammh{\Hl}{aa'}\ \psi_{a'}^{b'}(\pl)\
  \gdagg{\Hl}{b'b} . 
\eqb
Note that the blockspin transformation is diagonal with respect to spin and
flavor for continuum momenta within the first Brillouin zone $\Bc$, 
yet not for all $l \neq 0$.

We are now prepared to compute the perfect action for a RGT of the 
Gaussian type. Starting from a continuum action with a general
mass term $m_{b}$ -- which does not need to be flavor degenerate --
the perfect lattice action $S[\bar \phi ,\phi ]$ is defined as
\Eqa
    &   & e^{-S[\phib,\phi]}  
    \ = \ \int\!\Dpsi \int\!\Deta \ 
\ \exp \Bigl\{ -\int \!\! \frac{dq}{N_f(2\pi)^d} \ \psib_a^b(-q) 
\left( i\gmu_{aa'}q_\mu+m_b \right) \psi_{a'}^b(q) \Bigr\} \nonumber\\
&\times & \exp \Bigl\{ \intpsl \ \Bigl[ \ 
[ \phib_a^b(-p) \, - \, \sqrt{N_f} \suml
\Pi(\pl)\ \psibh_a{}^b(-p-2\pi l)\ ]\ \eta_a^b(p) \nonumber\\
& & + \etab_a^b(-p) \ [ \ \phi_a^b(p) \, - \, 
\sqrt{N_f} \suml \Pi(\pl)\ \psih_a{}^b(\pl)\ ] \nonumber \\
& & + \etab_a^b(-p) D_{aa'}^{b}(p) \eta_{a'}^{b}(p) \Bigr] \Bigr\}
\label{BST}
\Eqb
A non-zero term 
\eqa
D_{aa'}^{b}(p) \ = \ \gmu_{aa'}D_\mu^b(p) \ + \ \delta_{aa'} \, D_0^b(p)
\eqb
``smears out'' the blockspin transformation, as in Ref. \cite{BBCW}. 
This term is used to optimize locality of the resulting perfect
action; it will be specified later on.
The Gaussian integrals over $\psi$,$\psib$ and $\etab$,$\eta$ can be evaluated
by substitution of the classical fields leading to
\footnote{We ignore constant factors in the partition function.}
\eqa \label{Sp-result}
    S[\phib,\phi] \ = \ \intpsl \ \Bigl[ \ \phib_a^b(-p) \
                        G^{-1}{}_{aa'}^{bb'}(p) \ \phi_{a'}^{b'}(p) \ \Bigr] \ ,
\eqb
with the lattice propagator 
\eqa \label{propagator}
    G_{aa'}^{bb'}(p) \, = \, D_{aa'}^{bb'}(p)  
    \, + \, \suml \Bigl( \Pi(\pl)^2 \, \gammh{\Hl}{bd} \ \frac{[-i(-1)^{k_\nu}
            \gammh{\mu}{aa'} (p+2 \pi l)_\mu + m_d] \, \delta_{dd'}}
            {(p+2 \pi l)^2+m_d^2} \gdagg{\Hl}{d'b'} \Bigr) \ .  
\eqb
Note that $G$ is flavor diagonal, $G^{bb'}=G^{b}\delta^{bb'}$, because
$\gamma^{\Hl}\gK{}\gdagg{\Hl}{}$ is diagonal iff $\gK{}$ is. 
In particular, for a degenerate mass term the adjungation with $\gamma^{\Hl}$
is trivial, and the lattice propagator is proportional to $\delta_{bb'}$ in
flavor space. We define
\eqa \label{def-Qb} 
    G^b_{aa'}(p) \ = \ -i \sum_\mu \gmu_{aa'} \, Q^b_\mu(p)
                 \ + \ \delta_{aa'} Q^b_0(p) \ ,   
\eqb
hence $Q_\mu^b,Q_0^b$ become
\Eqa
         Q^b_\mu(p) 
   & = & D_\mu^b(p) \, + \suml \Pi(\pl)^2  
         \frac{1}{N_f} \sum_{b'} \sum_K \epsilon_K(l)\, \gK{bb}\gstar{K}{b'b'}
         \frac{(-1)^{l_\mu}(\pl)_\mu}{(\pl)^2+m_{b'}^2}\,,\qquad \label{Qmub}\\
         Q^b_0(p) 
   & = & D_0^b(p) \, + \suml \Pi(\pl)^2 
         \frac{1}{N_f} \sum_{b'} \sum_K \epsilon_K(l)\, \gK{bb}\gstar{K}{b'b'}
         \frac{m_{b'}}{(\pl)^2+m_{b'}^2}  \, .         \qquad \label{Q0b}
\Eqb 
In case of a non-degenerate mass $m_b$ the sums $\sum_{b'}\sum_K$ can not be
contracted according to \eqref{sumH}, due to the sign factor
$\epsilon_K(l)=\epsKtext$. However,
in the degenerate case $m_b=m$, with flavor independent smearing terms
$D^{b}=D$, we simply obtain
\Eqa 
    Q_\mu(p) & = & D_\mu(p) \ + \ \suml \Pi(\pl)^2 \, \frac{ (-1)^{l_\mu}
                   (\pl)_\mu}{(\pl)^2+m^2} \ ,             \label{Qmub-deg}\\
    Q_0(p)   & = & D_0(p) \ + \ \suml \Pi(\pl)^2  \frac{m}{(\pl)^2+m^2} \ .
                                                           \label{Q0b-deg}
\Eqb 

The perfect action in real space arises from \eqref{Sp-result} inserting the
momentum representation \eqref{ytopab}. After some $\gamma$--matrix algebra
\cite{Dilg}, we arrive at
\Eqa
S[\phib,\phi] & = & \sum_{y,y'} \phib(y) \, m(y,y') \, \phi(y') \ , 
                                                         \label{SlattK1}\\
m(y,y') & = & \sum_K \rho^K(y')\rho(y-y',y') \ M^K(y-y')\ . \label{SlattK2}
\Eqb
Corresponding to a lattice propagator diagonal in flavor space, the sum over
$K$ runs over multi-indices $K\in\Dc$ with diagonal $\gK{}$, in the
Weyl basis $\Dc \, = \, \{\emptyset,\ 1\,2,\ 3\,4,\ 1\,2\,3\,4\}$.
The sign factors $\rho(z,y)\equiv\rho(H(z),H(y))$ arise from 
$\gH{} \gK{} = \rho(H,K) \, \gamma^{H\Delta K}$, 
where $H\Delta K=(H\cup K)\backslash (H\cap K)$, 
and $\rho^K(y)$ is given by $\rho(H(y),K) \rho(K,H(y))$.
By symmetry, the only non-zero contributions to $M^{K}(y)$ are
\Eqa
    M^K_\mu(y) & = & i \rho(\mu,K) \intpc \ e^{-ipy} \ M^K_\mu(p)
                     \qquad \mbox{ for } H(y)=\mu\Delta K\ ,\qquad\label{MKy1}\\
    M^K_0(y)   & = & \ \qquad\qquad \intpc \ e^{-ipy} \ M^K_0(p) 
                     \qquad \mbox{ for } H(y)=K        \ ,  \qquad\label{MKy2}\\
    \mbox{with} \quad M^K_{\mu,0}(p) 
               & = & \frac{1}{N_f} \sum_b\gstarK{bb}\,\frac{Q_{\mu,0}^b(p)}
                     {\sum_\mu Q_\mu^b(p)^2 + Q_0^b(p)^2}\ .\qquad\label{def-mK}
\Eqb
The flavor degenerate case leads to vanishing components for $K\neq\emptyset$,
and for $M^\emptyset_{\mu,0} \equiv M_{\mu,0}$ we simply obtain
\eqa
    M_{\mu,0}(p) \ = \ \frac{Q_{\mu,0}(p)}
    {\sum_\nu Q_\nu(p)^2 + Q_0(p)^2} \ . \label{def-mK-deg}
\eqb

It has been proven in Ref. \cite{Dilg} that the couplings given by the fermion
matrix $m(y,y')$ are local, i.e.\ they decay faster than any power of
$|y-y'|$. 
For that, certain periodicity properties apply, which translate into (for
simplicity of notion let $\mut$ denote either $\mu$ or $0$, $K\Delta0\equiv K$)
\eqa \label{periodicity}
    Q^b_{\mut}(p) \ = \ \sum_{K\in\Dc} \gK{bb} \ Q^K_{\mut}(p) \ , \quad
    Q^K_{\mut}(p+2\pi\hat{\nu}) \ = \ 
    \twocase{ -Q^K_{\mut}(p) }{ \nu \,     \in \ K\Delta\mut }
            {  Q^K_{\mut}(p) }{ \nu \, \not\in \ K\Delta\mut } \ .
\eqb
Again, we sum over diagonal $\gamma$--matrices only. It is provided that
the corresponding requirements are met for the smearing terms $D^b_{\mut(p)}$
within $Q^b_{\mut}(p)$, see below. 
In consequence, the fermion matrix components $M^K_{\mut}(p)$ obey periodicity
conditions analogous to $Q^K_{\mut}(p)$ and the integrands of 
\eqRef{MKy1}{MKy2}
are periodic with respect to the Brillouin zone $\Bc$ and analytic in a
strip around the real axis. This implies locality of the perfect action.

The coupling of even and odd lattice points is due to 
the $M^K_\mu$ components of
the fermion matrix; the $M^K_0$ components couple even--even and odd--odd. 
We add without proof that even-odd decoupling of the Hermitian matrix
$m^\dagger\,m$ can be shown in any even dimension $d$ with arbitrary
(non-degenerate) mass terms for truncated versions of the
perfect fermion matrix $m(y,y')$, see Ref. \cite{Dilg} for $d=2$. 
This is a useful property in simulations
with Hybrid Monte Carlo algorithms. However, with (non-perfect) coupling to a
gauge field and non-zero mass term (i.e.\ with even--even, odd--odd as well as
even--odd couplings), this is not true in general. Yet, since even-odd
decoupling of $m^\dagger\,m$ is a perfect property, it might be imposed as a
construction requirement for a (quasi) perfectly gauged fermion vertex.

%%%%%%%%%%%%%%%%%%%%%%%%%%%%%%%%%%%%%%%%%%%%%%%%%%%%%%%%%%%%%%%%%%%%%%%%%%%
\section{Optimization of Locality} \label{Locality}

In the blocking scheme described so far, there is quite some freedom
left. In particular, 
we may use averaging functions different from 
%the standard block average given by 
$\Pi_{BA}$, % in \secref{Scheme}
and we can choose the smearing term $D$ in \eqref{BST}. 
In both cases we aim at optimization of the locality in the resulting 
perfect action. 

Let us first discuss the averaging scheme. In \eqref{BST} we implicitly assumed
the same blocking of $\psi$ and $\psib$, given by the weight function
$\Pi(x)$ resp. its Fourier transform $\Pi(p)$. Now we consider the case of a
block average $\Pi=\Pi_{BA}$ for $\psi$ ($\psib$) only, while $\psib$ 
($\psi$) is put on the lattice by decimation, $\Pi(x)=\delta
(x)$. Thus we obtain a single factor $\Pi(\pl)$ in \eqref{propagator} with
\eqa
    \Pi(p)  \ = \ \Pi_{BA}(p) \ = \ \prod_\mu \frac{\ph_\mu}{p_\mu} \ , \qquad
    \ph_\mu \ = \ 2\sin \frac{p_{\mu}}{2} \ .
\eqb
In case of a $\delta$--blockspin transformation ($D=0$), this 
means to identify the averaged continuum and lattice 2-point functions
\eqa \label{twopoint}
    \langle \ \phi(y) \ \phib(y') \ \rangle \ = \ 
    \int_{[y]}dx\ \langle \ \varphi(x,H(y)) \ \varphib(y',H(y')) \ \rangle \ .
\eqb
Due to translation invariance it doesn't matter whether we average source or
sink of the continuum expression, or whether we allocate the space directions
to be integrated over to source and sink in some way. The last point of view 
may be used to make a closer contact to the construction of staggered fermions
from DK fermions in the continuum \cite{BJ}, as discussed in Ref. \cite{Dilg}.
We call this blocking scheme {\em partial decimation}. For
the 2-point functions every space direction is integrated over once; 
therefore we do not run into the difficulties arising for
blockspin transformations with 
complete decimation, which do not have a corresponding perfect action.
%which would not even be well-defined for blocking from the
%continuum, i.e.\ with infinite blocking factor.

For both blocking schemes, block average for $\psi$ and $\psib$ (BA) and
partial decimation (PD), we now want to optimize locality of the
couplings by making use of the smearing terms in \eqRef{Qmub}{Q0b}.
As an optimization criterion it has been suggested to require
that in the effectively 1d case -- with momenta
$p=(p_{1},0,\dots ,0)$ -- the couplings are restricted to nearest
neighbors as in the standard action \cite{QuaGlu,Dallas}.%
\footnote{It has been shown in Ref. \cite{scal} that this 
criterion does optimize locality in $d=4$ for scalar particles 
over a wide range of masses.}
In the degenerate case we require
\eqa
   i\gamma_1 \, M_{1}(p_{1},0,\dots ,0) \ + \ M_{0}(p_{1},0,\dots ,0) \ = \  
   f(m) \, [ \, i \hat p_{1} \, \gamma_{1} \ + \ \hat m \, ] , \label{requ}
\eqb
with $\hat m \vert_{m=0}=0$ and $f(0)=1$.
Our ansatz for the Gaussian smearing term reads
\begin{equation}
D_{\mu}(p) = c(m)\hat p_{\mu} , \qquad D_{0}(p) = a(m).
\end{equation}
Requirement (\ref{requ}) can be fulfilled in both blocking schemes
we are considering, if we specify the RGT as follows
\begin{eqnarray}
c_{BA}(m) & = & [\cosh (m/2)-1)] \, / \, m^2 \ , \quad
a_{BA}(m) \ = \ [\sinh(m) -m]    \, / \, m^2 \ , \nonumber\\ 
c_{PD}(m) & = & 0 \ , \qquad \qquad \qquad \qquad \qquad \!
a_{PD}(m) \ = \ [\cosh (m/2)-1] \, / \, m   \ . \label{locop}
\end{eqnarray}
In both cases, we obtain $\hat m = 2 \sinh (m/2)$.
For $m=0$ a non-vanishing static smearing term $a(0)$
would explicitly break the remnant chiral symmetry in the
fixed point action. Therefore, the static term should vanish 
for optimized locality, as it does in both cases.
Furthermore, in the PD scheme the chiral limit is optimized 
for locality by
a simple $\delta$ function RGT. As an advantage of this property --
which is not provided by the BA scheme -- there is a direct relation
between the $n$-point functions in the continuum and on the
lattice. In addition, the extension to interacting theories
might involve numerical RGT steps in the classical limit, 
which also simplify in the absence of a Gaussian smearing term.
Finally, a non-vanishing term $c(m)$ causes complications
in the inclusion of a gauge interaction; the PD scheme avoids
such problems.

The decay of the couplings $m(x,0) = m(x)$
in the massless case for $d=2$ and $d=4$ is
shown in Figures 1 and 2.
%figalt - Anfang
%%%%%%%%%%%%%%%%%%%%%%%%%%%%%%%%%%%%%%%%%%%%%%%%%%%%%%%%%%%%%%%%%%%%
%% - included  ps-figure --- 2 in one row
%\begin{figure}[htb]
%\begin{center}
%\epsfig{file=coupl0-0-2d.ps,width=0.5\textwidth}
%\epsfig{file=coupl0-0-4d.ps,width=0.5\textwidth}
%\parbox{\textwidth}{ \caption{\label{coupl00} \sl 
%The decay of couplings in (1,0) direction for $d=2$ (left) and $d=4$ (right) 
%in the massless case: BA without smearing (diamonds), BA optimized (squares), 
%PD (triangles).
%}}
%\end{center}
%\end{figure}
%figalt - Ende
%%%%%%%%%%%%%%%%%%%%%%%%%%%%%%%%%%%%%%%%%%%%%%%%%%%%%%%%%%%%%%%%%%%%
% - included  ps-figure -------------------------
\begin{figure}[htb]
\begin{center}
\epsfig{file=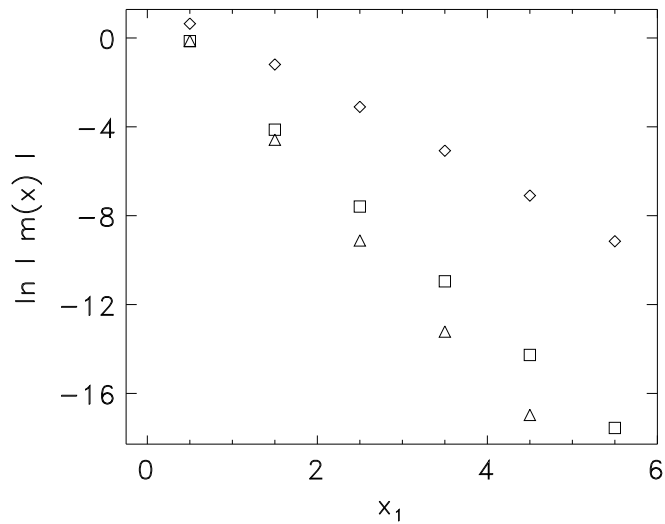,width=\figwidth}
\parbox{\textwidth}{ \caption{\label{coupl00-2d} \sl 
The decay of couplings in (1,0) direction for $d=2$ in the massless case: 
BA without smearing (diamonds), BA optimized (squares), PD (triangles).
}}
\end{center}
\end{figure}
%%%%%%%%%%%%%%%%%%%%%%%%%%%%%%%%%%%%%%%%%%%%%%%%%%%%%%%%%%%%%%%%%%%%
% - included  ps-figure -------------------------
\begin{figure}[htb]
\begin{center}
\epsfig{file=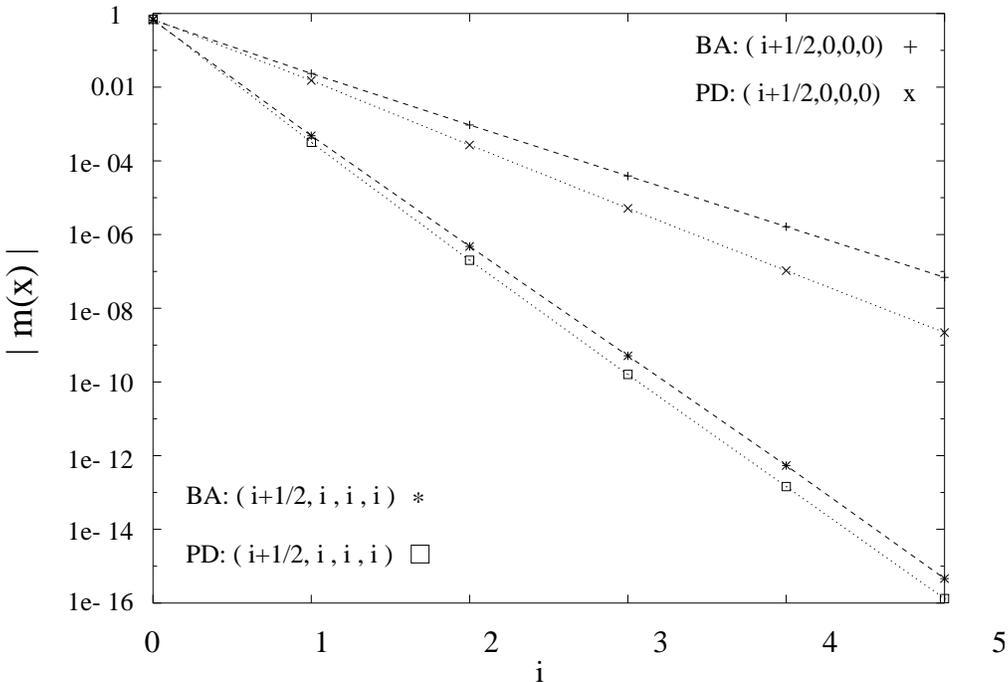,width=\figwidth,angle=270}
\parbox{\textwidth}{ \caption{\label{coupl00-4d} \sl 
The decay of 4d couplings in various directions in the massless case: 
BA with optimal smearing term and PD. We see that the latter
decays faster.
}}
\end{center}
\end{figure}
%%%%%%%%%%%%%%%%%%%%%%%%%%%%%%%%%%%%%%%%%%%%%%%%%%%%%%%%%%%%%%%%%%%%
We see that the PD blocking scheme works better. In the 
non-degenerate case the simplest ansatz for a smearing term is 
(corresponding to
the staggered fermion action with non-degenerate mass \cite{non-deg})
\eqa
    D^b_\mu(p) \ = \ c \, \ph_\mu \ , \qquad 
    D^b_0(p)\ =\ \sum_{K\in\Dc} \gK{bb}\ a_K \prod_{\mu\in K} \cos(p_\mu/2)\ .
\eqb
In this case, we obtained -- by means of numerical optimization -- a similar 
decay of couplings as with degenerate masses, yet no strict 1-dimensional
ultralocality. For $d=2$ it appears that a non-degenerate parameter 
$a_{12} \neq 0$ does not improve the coupling decay significantly, so we
worked with $a_{12}=0$. Again the PD scheme, where we assumed $c_{PD}=0$, led
to a more local action, see Figure 3.
%%%%%%%%%%%%%%%%%%%%%%%%%%%%%%%%%%%%%%%%%%%%%%%%%%%%%%%%%%%%%%%%%%%%
% - included  ps-figure -------------------------
\begin{figure}[htb]
\begin{center}
\epsfig{file=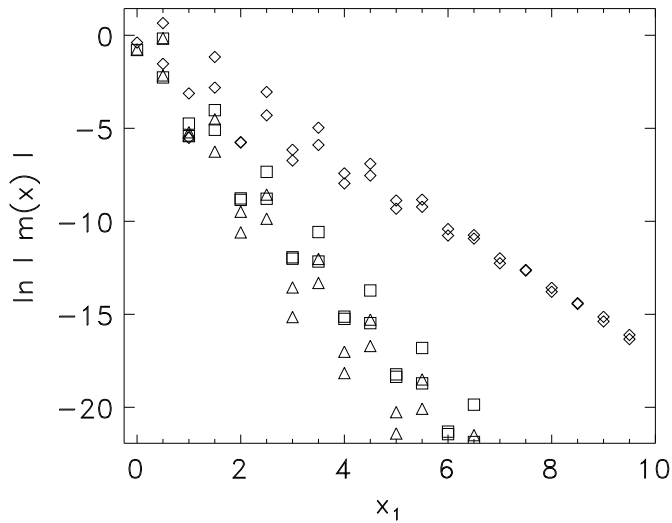,width=\figwidth}
\parbox{\textwidth}{ \caption{\label{coupl01-2d} \sl 
The decay of couplings in (1,0) direction for $d=2$ with masses
$m_1\!=\!0,\,m_2\!=\!1$: BA without smearing (diamonds), 
BA optimized (squares), PD (triangles).
}}
\end{center}
\end{figure}
%%%%%%%%%%%%%%%%%%%%%%%%%%%%%%%%%%%%%%%%%%%%%%%%%%%%%%%%%%%%%%%%%%%%

%%%%%%%%%%%%%%%%%%%%%%%%%%%%%%%%%%%%%%%%%%%%%%%%%%%%%%%%%%%%%%%%%%%%%%%%%%%
\section{Truncation Effects} \label{Truncation}
The litmus test of any blocking scheme is given by the truncation effects for
a practicable number of remaining couplings. These effects are minimized by
maximal locality. Yet, the truncation scheme itself may have some impact on
the truncation errors. An elegant procedure has been proposed in 
Ref. \cite{LAT96} for Wilson fermions.
First the perfect action is constructed  on a small lattice
volume $N^{d}$ by restricting the momentum components to the discrete
values $p_{\mu} = 2\pi n_{\mu}/N \in ]-\pi ,\pi ]$, $n_{\mu} \in \Z$.
Typically, one chooses $N=3$, and the resulting couplings 
are then used also in a large volume too, 
where they are not exactly perfect any more.
This truncation scheme has the virtues of automatically correct
normalizations, and a simplification of the 
numerical evaluation as opposed to a truncation in coordinate space.
Furthermore, the mapping to the corresponding truncated perfect
action in a lower dimension remains exact; for instance, one
can reproduce the effectively 1d nearest neighbor action starting
from $d>1$ by summing over the extra dimensions, which provides
a sensitive test of the numerical accuracy.
Provided a good locality, this method works well for Wilson-type
fermions, gauge fields \cite{LAT96} and scalars \cite{scal}.

For staggered fermions, only a crude truncation in coordinate 
space has been applied so far \cite{BBCW}. 
One includes the coupling distances $\pm 1/2,\ \pm 3/2$ in the $\mu$ direction,
and $1,0,-1$ in the non-$\mu$ directions. This involves
more couplings than the $N=3$ truncated Wilson-type fermion
(called ``hypercube fermion''), whereas the number of degrees of freedom
is the same in both cases.
% hd- Nevertheless, the quality of its spectral
% and thermodynamic properties did not reach the level of the
% ``hypercube fermion'', although it was improved over the standard
% staggered formulation. Here we try to achieve such a high
% quality also for a truncated perfect staggered fermion,
% based on increased locality and also on an improved 
% truncation scheme, closer to the elegant procedure described above.
However, although significant improvement has been achieved compared to the
standard staggered formulation, the quality of its spectral
and thermodynamic properties did not reach the level of the
``hypercube fermion''. 
In order to arrive at results of similar quality, 
we adapt the above truncation 
procedure -- together with the PD scheme -- to the case of staggered fermions.

\begin{figure}[htb]
\begin{center}
\epsfig{file=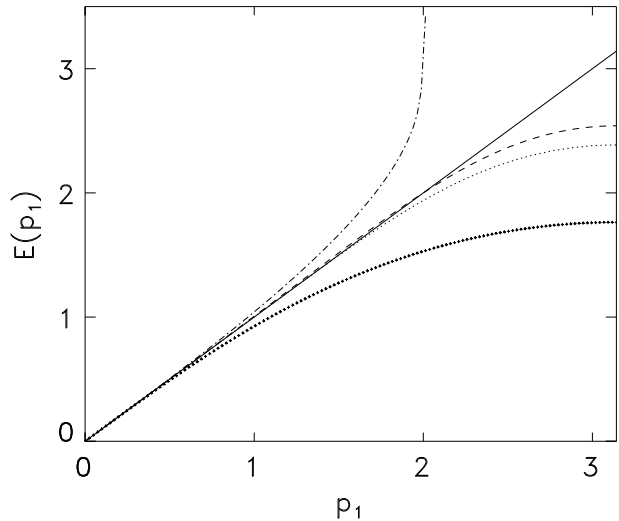,width=\figwidth}
\parbox{\textwidth}{ \caption{\label{spectrum2d-00} \sl 
The massless $d=2$ spectrum. We compare standard staggered (fat dots) and 
perfect spectrum (full line) with perfect actions truncated  with $N=3$ 
and mixed periodic boundary conditions: block average 
(dotted), partial decimation (dashed). The last case is also 
plotted for truncation with periodic boundary conditions (dashed-dotted).
}}
\end{center}
\end{figure}

We treat the components of the fermion matrix $M^K_{\mut}(p),\mut=\mu,0$
separately, as they show different behavior under reflections, 
%$\Pi^\nu: y_\nu,p_\nu \rightarrow -y_\nu,-p_\nu$
\eqa \label{reflections}
M^K_{\mut}(p_{1}\dots p_{\nu -1},-p_{\nu},p_{\nu+1} \dots p_{d}) \ = \ 
    \twocase{ -M^K_{\mut}(p) }{ \nu \,     \in \ K\Delta\mut }
            {  M^K_{\mut}(p) }{ \nu \, \not\in \ K\Delta\mut } \ .
\eqb 
Again we use the short-hand notation $K\Delta\mut=K$ for $\mut=0$. 
Truncation is achieved by discrete Fourier transformation and a discrete
support given by $c_N(y_\mu)=1,1/2,0$ for $|2y_\mu|<,=,>N$, respectively,
\eqa
    M^K_{\mut;N}(y) \ = \ \prod_\mu \frac{2\pi c_N(y_\mu)}{N} \
    \sum_{p \in \Bc^K_{\mut;N}} e^{-ipy} M^K_{\mut}(p) \ .
\eqb
It is the set of discrete momenta $\Bc^K_{\mut;N}$ which depends on the
reflection properties given by $\mut,K$. We choose partially antiperiodic
boundary conditions in $y$--space,
\eqa
    p \in \Bc^K_{\mut;N} \ \Leftrightarrow \ 
    p_\nu = \twocase{(2n+1)\pi/N}{\nu \,     \in \ K\Delta\mut}
                    {    2n\pi/N}{\nu \, \not\in \ K\Delta\mut}, \ n\in\Z,\ \
    p_\nu \in ]-\pi,\pi ] \ .
\eqb
It is easily verified that the transformation of the truncated components 
back to momentum space reproduces the perfect values at 
$p\in\Bc^K_{\mut;N}$ ,
\eqa
    M^K_{\mut;N}(p) \equiv \sum_y e^{ipy}\,M^K_{\mut;N}(y) \ = \  M^K_{\mut}(p)
    \ \mbox{ for } \ p \in \Bc^K_{\mut;N} \ . 
\eqb 
The components $M^K_{\mut}(y)$ inherit the reflection behavior of
\eqref{reflections} by Fourier transformation. Therefore, treating them as
periodic functions in $y$--space by discrete momenta $p_\nu=2\pi n/N,n\in\Z$ 
would make them vanish at the boundaries $y_\nu = \pm N/2$ for 
$\nu\in K\Delta\mut$ artificially. We avoid this effect by the
above choice of discrete momenta. It pays off
by a drastic reduction of truncation 
effects, see \fig{spectrum2d-00} for the 2d spectrum in the 
massless case using partial decimation. 
%%%%%%%%%%%%%%%%%%%%%%%%%%%%%%%%%%%%%%%%%%%%%%%%%%%%%%%%%%%%%%%%%%%%
%% - included  ps-figure -------------------------
%\begin{figure}[htb]
%\begin{center}
%\epsfig{file=spec0-0ext.ps,width=0.5\textwidth}
%\epsfig{file=spec0-1.ps,width=0.5\textwidth}
%\parbox{16cm}{ \caption{\label{spectrum2d} \sl 
%The $d=2$ spectrum for $m_{1,2}=0$ and $m_1=0, m_2=1$. We compare standard 
%staggered (crosses) and perfect spectrum (full line) with effective actions 
%truncated  with $N=3$ and partially antisymmetric boundary conditions: block 
%average (dashed--dotted), partial decimation (dashed). The last case is also
%plotted for truncation with periodic boundary conditions (dotted).
%}}
%\end{center}
%\end{figure}
%figalt - Ende
%%%%%%%%%%%%%%%%%%%%%%%%%%%%%%%%%%%%%%%%%%%%%%%%%%%%%%%%%%%%%%%%%%%%
% - included  ps-figure -------------------------
%%%%%%%%%%%%%%%%%%%%%%%%%%%%%%%%%%%%%%%%%%%%%%%%%%%%%%%%%%%%%%%%%%%%
% - included  ps-figure -------------------------
\begin{figure}[htb]
\begin{center}
\epsfig{file=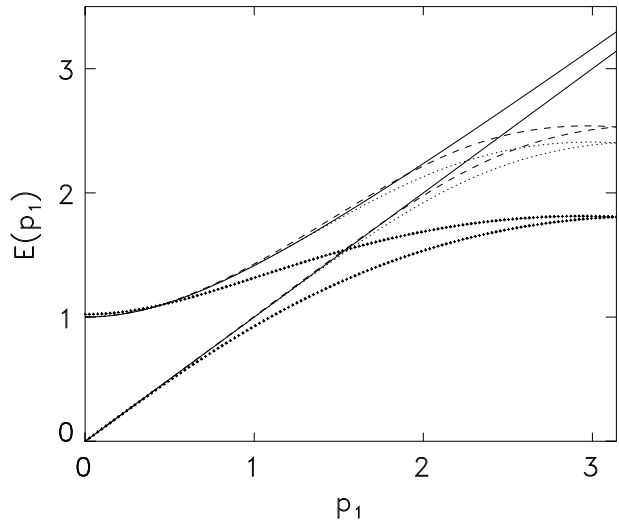,width=\figwidth}
\parbox{\textwidth}{ \caption{\label{spectrum2d-01} \sl 
The $d=2$ spectrum for $m_1=0, m_2=1$. 
Standard staggered (fat dots) and perfect
spectrum (full line) compared with perfect actions truncated  with $N=3$ and 
mixed periodic boundary conditions: block average (dotted), 
partial decimation (dashed). 
}}
\end{center}
\end{figure}
%%%%%%%%%%%%%%%%%%%%%%%%%%%%%%%%%%%%%%%%%%%%%%%%%%%%%%%%%%%%%%%%%%%%
The values of the perfect couplings, truncated by {\em mixed
periodic boundary conditions} as described above, are given in Table
\ref{tabcop}.
\begin{table}
\begin{center}
\begin{tabular}{|c|c|c|c|c|c|}
\hline
$d=4$ & & & $d=2$ & & \\
\hline
& PD & BA & & PD & BA \\
\hline
\hline
(0.5,0,0,0) & {\bf 0.696388} &  0.663116 & (0.5,0) & {\bf 0.878234} 
&  0.866299 \\
\hline
(0.5,1,0,0) & {\bf 0.040981} &  0.045925 & (0.5,1) & {\bf 0.060883} 
&  0.066850 \\
\hline
(0.5,1,1,0) & {\bf 0.004481} &  0.004871 & & & \\
\hline
(0.5,1,1,1) & {\bf 0.000495} &  0.000361 & & & \\
\hline
(1.5,0,0,0) & {\bf 0.015219} &  0.023441 & (1.5,0) & {\bf 0.010425} 
&  0.016043 \\
\hline
(1.5,1,0,0) &{\bf -0.000431} & -0.000639 & (0.5,1) &{\bf -0.005212} 
& -0.008022 \\
\hline
(1.5,1,1,0) &{\bf -0.000767} & -0.001211 & & & \\
\hline
(1.5,1,1,1) &{\bf -0.000428} & -0.000635 & & & \\
\hline
\hline
\end{tabular}
\end{center}
\caption{The couplings of the perfect action for massless staggered
fermions, constructed from the ``partial decimation'' (most successful)
and from the optimized ``block average'' scheme,
and truncated by mixed periodic boundary conditions.
The couplings are odd in the half-integer
component, and even in all other components. Among the latter
there is also permutation symmetry.}
\label{tabcop}
\end{table}

Furthermore, the truncation effects are significantly stronger for 
even truncation distances $N$; in particular, for $N=2$ and $N=4$
the energies become complex-valued at large momenta.
 
For $N=3$ -- which we consider to be tractable in numerical
simulations -- the
amount of improvement is already striking. We compare the spectra for
$d$\,=\,2 in the PD and BA blocking scheme in the massless case
(\fig{spectrum2d-00}) and 
for non-degenerate masses $m_1=0, m_2=1$ (\fig{spectrum2d-01}).
The couplings derived from the PD blocking scheme turn out to be better,
in agreement with the higher degree of locality observed in the (untruncated)
perfect action.
%, see Fig.~(\ref{coupl00-2d},\ref{coupl01-2d}).
We see that the improvement is still good in the case of non-degenerate 
masses. In this case, we optimized the smearing parameters numerically,
as discussed in \secref{Locality}. The values are $a_{BA}=0.060$, 
$c_{BA}=0.125$, $a_{PD}=0.041$, $c_{PD}=0$.
The standard staggered fermion results for non-degenerate 
masses are calculated with the action proposed in Ref. \cite{non-deg}.
We emphasize that the spectra of (untruncated) perfect actions
are indeed perfect, i.e.\ identical to the continuum spectra
(up to the periodicity, which is inevitable on the lattice).
Hence the spectrum reveals directly the artifacts due to truncation.

%%%%%%%%%%%%%%%%%%%%%%%%%%%%%%%%%%%%%%%%%%%%%%%%%%%%%%%%%%%%%%%%%%%%%%%%% 
Figure 6 compares the massless spectra in $d=4$, 
and we see that the qualitative behavior observed in $d=2$ persists.
We compare the standard staggered fermion, 
and the optimized BA and PD fixed point fermions, 
both truncated by mixed periodic boundary conditions.
We see again that the PD scheme is superior.
Furthermore, we show for comparison also the dispersion relation
of a Symanzik improved action called ``p6'' from Ref. \cite{Bielef}. 
Symanzik improvement by additional couplings along the axes
(Naik fermion, \cite{Naik}) only yields a moderate quality
\cite{BBCW}, but the Bielefeld group suggests a number of
actions, where Symanzik improvement is achieved by diagonal
couplings. The p6 action is the best variant among them,
and also Figure 6 confirms its excellent level of improvement.
However, it is not obvious how this formulation can
include a general mass term in a subtle way.
\begin{figure}[htb]
\begin{center}
\epsfig{file=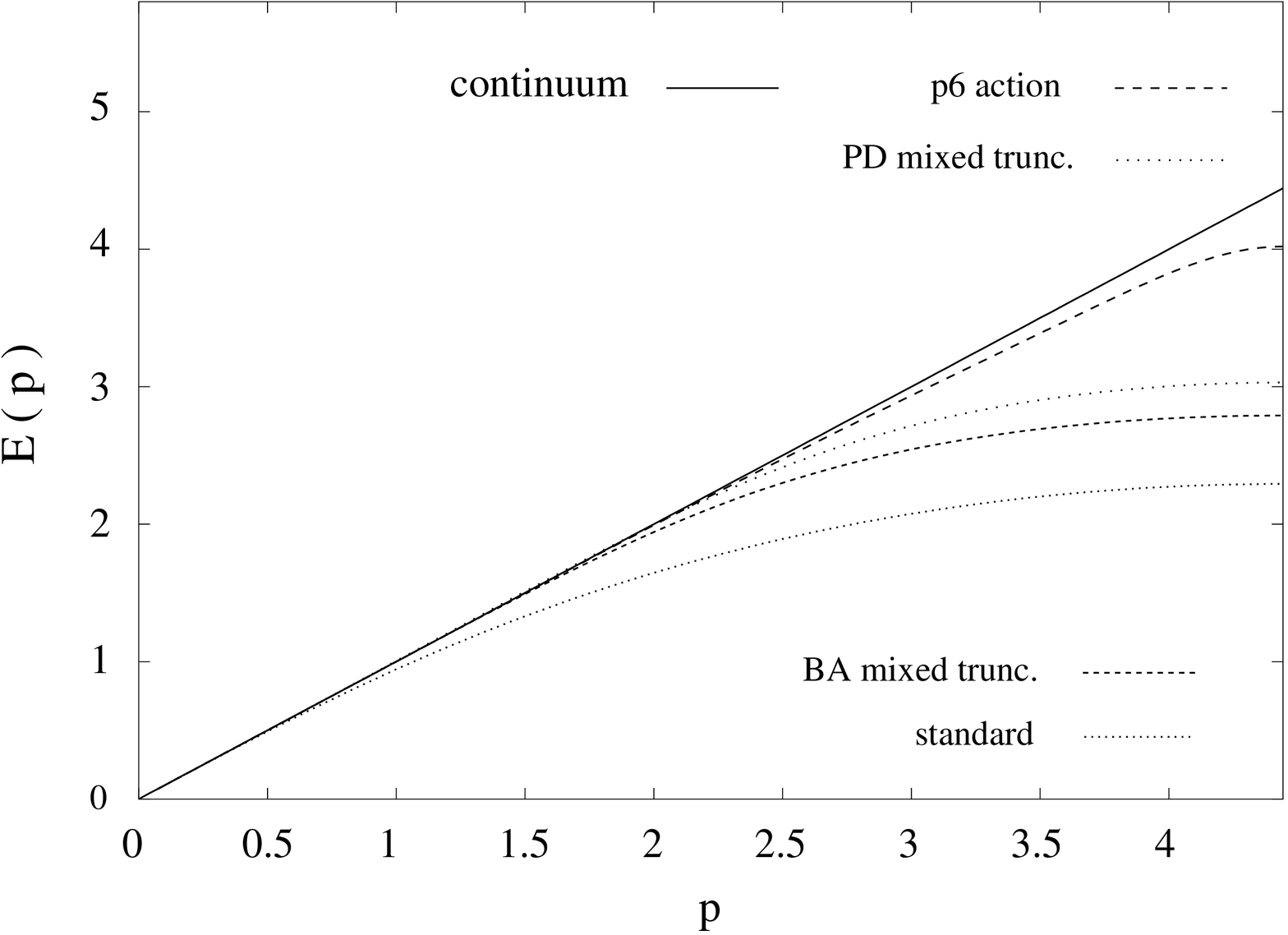,width=\figwidth,angle=270}
\parbox{\textwidth}{ \caption{\sl 
The dispersion relation for various types of 4d massless staggered
fermions along (1,1,0). We compare the standard action, 
a Symanzik improved action called ``p6'' from Ref. \cite{Bielef},
and the optimized BA and PD fixed point actions, truncated
with mixed periodic boundary conditions.}}
\label{spec4d}
\end{center}
\end{figure}

As a further test in $d=4$, we consider the thermodynamic
ratio $P/T^{4}$ ($P$: pressure, $T$: temperature). According
to the Stefan-Boltzmann law, this ratio is $7\pi^{2} /180$ 
for massless fermions in the
continuum, and a lattice action with many discrete points $N_{t}$
in the temporal direction will asymptotically reproduce that
value. However, the speed of convergence, and in particular
the behavior at small $N_{t}$, depend on the quality
of the action. In contrast to the spectrum, this ratio is not even 
exact for the fixed point action in \eqRef{Qmub-deg}{Q0b-deg}, because
of the ``constant factors'' that we ignored when performing the
functional integral in the RGT \eqref{BST}. 
Such factors may depend on the
temperature, so with respect to thermodynamics our action is not
fully renormalized \cite{LAT96}. However, it turns out that
the unknown factor is very close to 1, except for the regime
of very small $N_{t}$ (about $N_{t}\leq 3$), which corresponds to
extremely high temperature. So the main issue is again the 
contamination due to the truncation.

In Figure 7 we compare this thermodynamic scaling
for a variety of staggered fermion actions at $m=0$,
and again the PD scheme turns out to be very successful.
A similar level of improvement can be observed for the p6 action,
which is also here by far better than the Naik fermion. We conclude that
a good improvement method should in any case include diagonal
couplings, which are far more promising than additional couplings on
the axes (just consider rotational invariance, for example).
%Thus we can reduce the value of $N_{t}$, which is needed to see
%scaling setting in, 
Thus we can expand the range, in which a practically
accurate continuum behavior is observed,
by about a factor of 4 compared to standard staggered fermions
(see Figures 4, 5, 6 and 7).
%including also some
%improvements \`{a} la Symanzik, but the truncated perfect action
%in the PD scheme is again most successful. The quality of a Symanzik
%staggered fermion action, which is Symanzik improved by additional
%couplings on the axes (Naik fermion \cite{Naik}) is not rather
%moderate \cite{BBCW}, but an excellent level of improvement can also
%be achieved by Symanzik improvement involving diagonal couplings;
%as an example, we also show in Figure \ref{presfig} the so-called ``p6''
%action from Ref. \cite{Bielef}.
\begin{figure}[htb]
\begin{center}
\epsfig{file=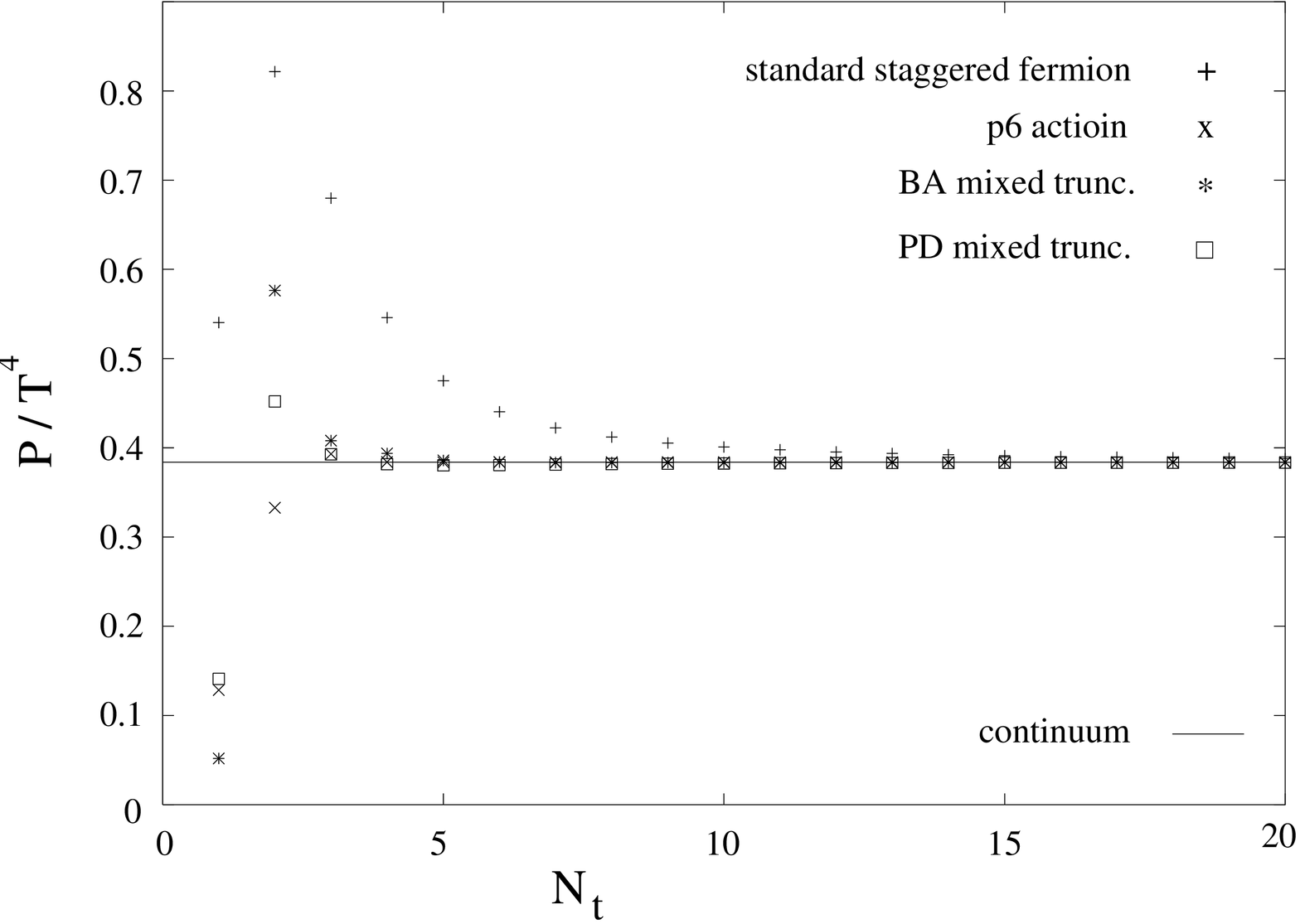,width=\figwidth,angle=270}
\parbox{\textwidth}{ \caption{\sl 
The Stefan-Boltzmann law for standard staggered fermions
and for the ``p6 action'', compared to
the truncated perfect fermions for the optimized BA and PD scheme.}}
\label{presfig}
\end{center}
\end{figure}

%%%%%%%%%%%%%%%%%%%%%%%%%%%%%%%%%%%%%%%%%%%%%%%%%%%%%%%%%%%%%%%%%%%%%%%%%%
\section{Conclusions}

Spectral and thermodynamic results for truncated perfect staggered
fermions have been presented before \cite{BBCW}, but the improvement
did not really reach a satisfactory level there. We could now
push that improvement significantly further, mainly thanks to
the new blocking scheme, which we call {\em partial decimation}, but 
also with the help of a new truncation technique ({\em mixed periodic}
boundary conditions). We now reached a level of excellent
improvement, similar to the results for truncated perfect
Wilson-type fermions.

As a novelty, we extended the construction of perfect actions
to non-degenerate flavors, and we could preserve the same level
of improvement after truncation also in that case.
This is potentially important 
%hd+ if one wants to study the decoupling of heavy flavors.
for the study of the decoupling of heavy flavors, or for QCD simulations 
with realistic quark masses.

At $m=0$, which is well described by
staggered fermions, the PD scheme is optimally local if we just use
a $\delta$ function RGT. This simplifies the relation
to the continuum $n$-point functions, and numerical RGT steps,
which could be performed for interacting theories (in the classical
limit).

The next step is the inclusion of gauge interactions, which is not
straightforward. However, also with this respect the absence
of a Gaussian smearing term in the chiral limit is of advantage,
because a momentum dependent smearing term -- 
as present for the block average
scheme -- causes complications in that step.

As a simple ansatz, one can insert standard link variables,
connect the coupled sites over the shortest lattice paths
and take the mean value over these paths.
This method is a simplification, which is not perfect, but for 
Wilson type fermions it leads for instance
to a drastically improved mesonic
dispersion relation \cite{LAT96}.
In that framework, applications to the charmonium 
spectrum \cite{Kostas}, pionic systems \cite{TdG},
and the suitable generalization of 
preconditioning techniques for the fermion matrix \cite{Norbert}
are under investigation.

We have implemented this method for truncated perfect staggered
fermions in $d=2$. Including also ``fat links'', we
obtained promising results for the scaling
properties in the Schwinger model \cite{prep} down to small 
inverse couplings, $\beta \leq 1$. 
In particular, the ``pion mass'' is drastically reduced compared
to the staggered standard action,
and the ``eta mass'' follows the prediction by asymptotic scaling very 
closely.\\

%\vspace{1cm}

\noindent
{\em Acknowledgment} \ We thank S. Chandrasekharan for his assistance.

\end{document}